\documentclass[preprint,aps,12pt,showpacs,nofootinbib,tightenlines]{revtex4}
\usepackage{amsmath}
\usepackage{amssymb}
\usepackage{epsfig}
\usepackage{graphicx}

\def \eff{{\text{eff}}}
\def\gl{\Gamma}
\def\d{{\rm d}}
\def\mh{\hat{m}}
\def\mvh{\mh_{D_s^*}}
\def\mlh{\mh_l}
\def\sh{\hat{s}}
\def\uh{{\hat{u}}}
\def\la{{\lambda}}

\def\be{\begin{equation}}
\def\ee{\end{equation}}
\def\ba{\begin{eqnarray}}
\def\ea{\end{eqnarray}}
\begin{document}

\title{The rare decay $B_c \rightarrow D_s^* \mu^+ \mu^-$ in a family non-universal $Z^{\prime}$ model}
\author{L\"u Lin-Xia$^1$} \email{lvlinxia@sina.com}
\author{Zhang Guo-Fang$^1$}
\author{Wang Shuai-Wei$^1$}
\author{Zhang Zhi-Qing$^2$}
\affiliation{1. Physics and electronic engineering college, Nanyang Normal University,
Nanyang, Henan 473061, P.R.China \\2. Department of Physics, Henan University of Technology,
Zhengzhou, Henan 450052, P.R.China }
\date{\today}

\begin{abstract}
Using the form factors calculated in the three-point QCD sum rules, we calculate the new physics contributions
to the physical observables of $B_c \to D_s^* \mu^+\mu^-$ decay in a family non-universal $Z^{\prime}$ model.
Under the consideration of three cases of the new physics parameters, we find that: (a) the $Z^{\prime}$ boson can
provide large contributions to the differential decay rates;
(b) the forward-backward asymmetry (FBA) can be increased by about $47\%$, $38\%$, and $110\%$
at most in S1, S2, and extreme limit values (ELV), respectively.
In addition, the zero crossing can be shifted in all the cases;
(c) when $\hat{s}>0.08$, the value of $P_L$ can be changed from $-1$ in the Standard Model (SM) to $-0.5$
in S1, $-0.6$ in S2, and $0$ in extreme limit values, respectively;
(d) the new physics corrections to $P_T$ will decrease the SM prediction
about $25\%$ for the cases of S1 and S2, $100\%$ for the case of ELV.

\vspace{1cm} \noindent {\bf Key words:} Rare decays, Standard Model, a family non-universal $Z^{\prime}$ model
\end{abstract}


\maketitle

%%%%%%%%%%%%%%%%%%%%%%%%%%%%%%%%%%%%%%%%%%%%%
\section{Introduction}
The Standard Model~(SM) of interactions among elementary particles is one of the best verified physics theories
up to now but there are many open fundamental questions remain unanswered within the scope of the SM.
High energy physics experiments are designed to address the open questions through the search of new physics~(NP)
using two complementary approaches. One is to discover the new particles at the high energy Large Hadron Collider (LHC).
The other is to search for the effects of NP through measurements of flavor physics reactions at lower energy scales
and evidence of a deviation from the SM prediction.

After the observation of the rare radiative decay $b \rightarrow s \gamma$~\cite{CLEO1995},
the flavor-changing neutral current~(FCNC) transitions became more attractive and since then
many works about rare radiative, leptonic and semileptonic decays have been intensively done
in the $B_{u,d,s}$ system~\cite{Ali2005}. Among these decays, semileptonic decay channels are
significant because their branching ratios are relatively larger. These works will be
more perfect if similar studies for $B_c$, observed in
1998 by CDF Collaboration~\cite{CDF1998}, are also included.

The charmed $B_c$ meson is a ground state of two heavy quarks $b$ and $c$.
Because of the two heavy quarks, the decays of the $B_c$ meson are
rather different from $B_u/B_d/B_s$ mesons.
Physicists therefore believe that the $B_c$ physics must be very rich
compared to the other $B$ mesons if the statistics reaches high level~\cite{CDF1998,Brambilla1,Brambilla2}.
At LHC, around $5 \times 10^{10}$ $B_c$ events per year are expected~\cite{Sun08,Altarelli08}.
The expected number events are motivating to work on the
$B_c$ phenomenology and this possibility will provide facilities to study the
observables of rare $B_c$ decays such as branching ratios, forward-backward asymmetry and polarization asymmetries.

The rare $B_c \rightarrow D_s^* l^+ l^-$ decays are proceeded by FCNC
transition of $b \rightarrow s l^+ l^-$, which are forbidden at the tree level in the SM,
and play an important role in the precision test of the SM. Meanwhile, they offer a valuable possibility
of an indirect search of NP for their sensitivity to the gauge structure
and new contributions. Up to now, the possible new physics contributions to $B_c \rightarrow D_s^* l^+ l^-$ decays have been
studied extensively, for example, by using model independent
effective Hamiltonian~\cite{Yilmaz2007},
in Supersymmetric models~\cite{AAhmed}, with fourth generation
effects~\cite{IAhmed}, and in single universal extra dimension~\cite{Yilmaz}.

When concentrating on the exclusive $B_c \rightarrow D_s^* l^+ l^-$ decays,
one needs to know the form factors. As for $B_c \rightarrow D_s^*$
transition, the form factors have been
calculated using different approaches, such as light front constituent quark
models~\cite{formfactor1}, a relativistic constituent quark model~\cite{formfactor2},
a relativistic quark model~\cite{formfactor3}, the Ward identities~\cite{formfactor4},
in light cone QCD~\cite{formfactor5,formfactor6}, and
QCD sum rules~\cite{formfactor7,formfactor8,formfactor9}.
In this work, we will adopt the form factors calculated in the three-point QCD sum rules~\cite{formfactor9}
to study the $Z^\prime$ effects on the observables for $B_c \rightarrow D_s^* \mu^+ \mu^-$ decay.

The general framework for non-universal $Z^{\prime}$ model has been developed in Ref.~\cite{Langacker}.
In this model, $Z^\prime$ gauge boson could be naturally derived by adding additional $U(1)^\prime$
gauge symmetry. Non-universal $Z^\prime$ couplings can induce FCNC $b\to s$ and $d$ transitions at tree level.
Its effects on $b \rightarrow s$ transition have received great attention and been
widely studied in the literature. The previous works in a family non-universal $Z^\prime$ model boson
redound to resolve many puzzles, such as "$\pi K$ puzzle"~\cite{Barger1,Chang1},
anomalous $\bar{B}_s-B_s$ mixing phase~\cite{Liu,Chang2} and mismatch
in $A_{FB}(B\to K^{\ast}\mu^{+}\mu^{-})$ spectrum at low $q^2$ region~\cite{Chang3,CDLv}.
Motivated by this, we will study the effects of the
$Z^\prime$ boson on the rare decay $B_c \rightarrow D_s^* \mu^+ \mu^-$.

This paper is organized as follows. In Section~2,
we present the effective Hamiltonian
responsible for the $b\to s l^+ l^-$ transition in both the SM and the family
non-universal $Z^{\prime}$ model. In this section we also present the matrix element,
and the expressions of various physical observables in $Z^{\prime}$ model. In Section~3, we show
the numerical results of the observables for $B_c \to D^*_s \mu^+\mu^-$ decay in the SM and
$Z^{\prime}$ model. The final section is the summary.

%%%%%%%%%%%%%%%%%%%%%%%%%%%%%%%%%%%%%%%%%%%%%%%%%%%%%%%%%%%%%%%%%%%%%%%%%%%%%
\section{Effective Hamiltonian, matrix elements and observables for $b\to s l^+ l^-$ decay}\label{theo}
%%%%%%%%%%%%%%%%%%%%%%%%%%%%%%%%%%%%%%%%%%%%%%%%%%%%%%%%%%%%%%%%%%%%%%%%%%%%%
At quark level, the rare semileptonic decay $b\to s l^+ l^-$ can be described in terms of
the effective Hamiltonian which is given by~\cite{Altmannshofer:2008dz,Chetyrkin:1996vx}
\begin{equation} \label{eq:Heff}
{\cal H}_{\eff} = - \frac{4\,G_F}{\sqrt{2}}V_{tb}V_{ts}^{\ast}
\sum_{i=1}^{10} C_i(\mu) O_i(\mu) \,.
\end{equation}
here the explicit expressions of $O_i$ can be found in Ref.~\cite{Altmannshofer:2008dz}, in which
\begin{eqnarray}\label{O910}
O_9=\frac{e^2}{g_s^2}(\bar{d}\gamma_\mu P_Lb)(\bar{l}\gamma^\mu l)\,,\quad O_{10}=\frac{e^2}{g_s^2}(\bar{d}\gamma_\mu P_Lb)(\bar{l}\gamma^\mu\gamma_5 l)\,.
\end{eqnarray}
The Wilson coefficients $C_i$ can be expanded perturbatively~\cite{Beneke:2001at,bobeth,bobeth02,Huber:2005ig}.
The effective coefficients $C_{7,9}^{eff}$, can be written as~\cite{Altmannshofer:2008dz}
\begin{eqnarray}\label{eq:effWC}
&& C_7^{\rm eff} = \frac{4\pi}{\alpha_s}\, C_7 -\frac{1}{3}\, C_3 -
\frac{4}{9}\, C_4 - \frac{20}{3}\, C_5\, -\frac{80}{9}\,C_6\,,
\nonumber\\
&& C_9^{\rm eff} = \frac{4\pi}{\alpha_s}\,C_9 + Y(\hat{s})\,, \nonumber\\
&& C_{10}^{\rm eff} = \frac{4\pi}{\alpha_s}\,C_{10}\,,
\end{eqnarray}
where the perturbative part $Y(q^2)$ stands for the matrix element of four-quark operators and is given by
\begin{eqnarray}
Y(q^2)&=&h(\hat{m_c},\hat{s})\big(\frac{4}{3}C_1+C_2+6C_3+60C_5\big)\,\nonumber\\
&&+\frac{1}{2}h(1,\hat{s})\big(-7C_3-\frac{4}{3}C_4-76C_5-\frac{64}{3}C_6\big)\,\nonumber\\
&&+\frac{1}{2}h(0,\hat{s})\big(-C_3-\frac{4}{3}C_4-16C_5-\frac{64}{3}C_6\big)\,\nonumber\\
&&+\frac{4}{3}C_3+\frac{64}{9}C_5+\frac{64}{27}C_6\,.
\end{eqnarray}
with $\hat{s}=q^2/m_{B_c}^2$, $\hat{m_c}=m_c/m_{B_c}$. We have neglected the resonance contribution.
For the detailed discussion of such resonance effects, we refer to Refs.~\cite{AAhmed,IAhmed,Yilmaz}.

Exclusive decay $B_c \to D^*_s \mu^+\mu^-$ is described in terms of matrix elements of the quark operators
in the effective Hamiltonian over meson states, which can be parameterized in terms of form factors.
The matrix elements of $B_c \to D^*_s$ transition are given by~\cite{Ali-prd61}
\ba \langle D_s^*(p) | (V-A)_\mu | B_c(p_{B_c})\rangle & = & -i
\epsilon^*_\mu (m_{B_c}+m_{D_s^*}) A_0(s) + i (p_{B_c}+p)_\mu (\epsilon^*p_{B_c})\,
\frac{A_+(s)}{m_{B_c}+m_{D_s^*}}\nonumber\\
\lefteqn{+ i q_\mu (\epsilon^* p_{B_c}) \,\frac{2m_{D_s^*}}{s}A_-(s)\,
+\epsilon_{\mu\nu\rho\sigma}\epsilon^{*\nu} p_{B_c}^\rho p^\sigma\,
\frac{2A_V(s)}{m_{B_c}+m_{D_s^*}}\,.}\hspace*{2cm}\label{eq:ff3} \ea
and
\ba \langle D_s^*(p) | \bar s \sigma_{\mu\nu} q^\nu (1+\gamma_5) b
| {B_c}(p_{B_c})\rangle & = & i\epsilon_{\mu\nu\rho\sigma} \epsilon^{*\nu}
p_{B_c}^\rho p^\sigma \, 2 T_1(s)\nonumber\\
& & {} + T_2(s) \left\{ \epsilon^*_\mu
  (m_{B_c}^2-m_{{D_s^*}}^2) - (\epsilon^* p_{B_c}) \,(p_{B_c}+p)_\mu \right\}\nonumber\\
& & {} + T_3(s) (\epsilon^* p_{B_c}) \left\{ q_\mu -
\frac{s}{m_{B_c}^2-m_{{D_s^*}}^2}\, (p_{B_c}+p)_\mu \right\}\label{eq:T} \ea
here $s=q^2$, $q_\mu=(p_{B_c}-p)_\mu$, and $\epsilon_\mu$ is polarization vector of the vector meson
$D_s^*$.

The dilepton invariant mass spectrum for $B_c \rightarrow D_s^* l^+ l^-$ decays can be expressed by~\cite{Ali-prd61,liwenjun}
\be
\frac{\d \gl}{\d\sh}  =
  \frac{G_F^2 \, \alpha^2 \, m_{B_c}^5}{2^{10} \pi^5}
      \left| V_{ts}^\ast  V_{tb} \right|^2 \, \uh(\sh) D
\ee
where the explicit expression of $D$ is
\ba
D& = &
\frac{|A|^2}{3} \sh \la (1+2 \frac{\mlh^2}{\sh}) +|E|^2 \sh
\frac{\uh(\sh)^2}{3}
        \Bigg.
        \nonumber \\
  & & + \Bigg. \frac{1}{4 \mvh^2} \left[
|B|^2 (\la-\frac{\uh(\sh)^2}{3} + 8 \mvh^2 (\sh+ 2 \mlh^2) )
          + |F|^2 (\la -\frac{ \uh(\sh)^2}{3} + 8 \mvh^2 (\sh- 4 \mlh^2))
\right]
        \Bigg.
        \nonumber \\
  & & +\Bigg.
   \frac{\la }{4 \mvh^2} \left[ |C|^2 (\la - \frac{\uh(\sh)^2}{3})
 + |G|^2 \left(\la -\frac{\uh(\sh)^2}{3}+4 \mlh^2(2+2 \mvh^2-\sh) \right)
\right]
        \Bigg.
        \nonumber \\
  & & - \Bigg.
   \frac{1}{2 \mvh^2}
\left[ {\rm Re}(BC^\ast) (\la -\frac{ \uh(\sh)^2}{3})(1 - \mvh^2 -
\sh)
\nonumber  \right. \Bigg.\\
& & + \left.  \Bigg.
       {\rm Re}(FG^\ast) ((\la -\frac{ \uh(\sh)^2}{3})(1 - \mvh^2 - \sh) +
4 \mlh^2 \la) \right]
        \Bigg.
        \nonumber \\
  & & - \Bigg.
 2 \frac{\mlh^2}{\mvh^2} \la  \left[ {\rm Re}(FH^\ast)-
 {\rm Re}(GH^\ast) (1-\mvh^2) \right] +\frac{\mlh^2}{\mvh^2} \sh \la |H|^2
   \,,
   \label{eq:dwbvll}
\ea
with $\mlh=m_l/m_{B_c}$, and $\mvh=m_{D_s^*}/m_{B_c}$. The kinematic variables $\hat{s}$ and $\hat{u}$
are the same as Ref.~\cite{Ali-prd61}. The auxiliary functions $A\,,B\,,C\,,E\,,F$ and $G$
which are combinations of the effective Wilson coefficients in Eq.~(\ref{eq:effWC}) and 
the form factors of $B_c \to D_s^*$ transition
can be found in Refs.~\cite{Ali-prd61,liwenjun}.
For the convenience of the reader, we present these functions in
the Appendix A.

The normalized forward-backward asymmetry (FBA) is defined as
\be {\cal A}_{FB}(\hat{s})=\int
d\hat{s}~\frac{\int^{+1}_{-1}dcos\theta\frac{d^2Br}{d\hat{s}dcos\theta}{\rm
Sign}(cos\theta)}
{\int^{+1}_{-1}dcos\theta\frac{d^2Br}{d\hat{s}dcos\theta}}.\ee
According to this definition, the explicit expression of FBA is:
\ba
\label{EqAFB}
\frac{d{\cal
A}_{FB}}{d\hat{s}}D &=&\hat{u}(\hat{s})
\hat{s} [Re(BE^{*}) +Re(AF^{*})] \,.
\ea

The lepton polarization can be defined as:
\be
\frac{d\Gamma(\hat{n})}{d\hat{s}}=\frac{1}{2}\big
(\frac{d\Gamma}{d\hat{s}}\big )_0[1
+(P_L\hat{e}_L+P_N\hat{e}_N+P_T\hat{e}_T)\cdot\hat{n}]
\ee
where the subscript $"0"$ stands for the unpolarized decay case.
$P_L$ and $P_T$ are the longitudinal and transverse polarization
asymmetries in the decay plane respectively, and $P_N$ is the
normal polarization asymmetry in the direction perpendicular to
the decay plane.

The lepton polarization asymmetry $P_i$ can be derived by
\be
P_i(\hat{s})=\frac{d\Gamma(\hat{n}=\hat{e}_i)/d\hat{s}-
d\Gamma(\hat{n}=-\hat{e}_i)/d\hat{s}}{d\Gamma(\hat{n}=\hat{e}_i)/d\hat{s}+
d\Gamma(\hat{n}=-\hat{e}_i)/d\hat{s}}\;
\ee
the results are
\ba P_L D&=&\sqrt{1-4\frac{\hat{m}^2_{l}}{\hat{s}}}\Bigg\{\frac{2\hat{s}\lambda}{3}
Re(AE^{*})+\frac{(\lambda+12\hat{s}\hat{m}^2_{D_s^*}
)}{3\hat{m}^2_{D_s^*}}Re(BF^{*})\Bigg.  \nonumber \\
&&\Bigg.-\frac{\lambda(1-\hat{m}^2_{D_s^*}-\hat{s})}{3\hat{m}^2_{D_s^*}}Re(BG^{*}+CF^{*})
+\frac{\lambda^2}{3\hat{m}_{D_s^*}}Re(CG^{*})\Bigg\},\label{plks}\\
P_N D&=&\frac{-\pi\sqrt{\hat{s}}\hat{u}(\hat{s})}{4\hat{m}_{D_s^*}}\Bigg\{
\frac{\hat{m}_l}{\hat{m}_{D_s^*}}\left[Im(FG^{*})(1+3\hat{m}^2_{D_s^*}-\hat{s})
\right.\Bigg. \nonumber \\
&&\Bigg.\left.+Im(FH^{*})(1-\hat{m}^2_{D_s^*}-\hat{s})-Im(GH^{*})\lambda
\right]\Bigg. \nonumber \\
&&\Bigg.+2\hat{m}_{D_s^*}\hat{m}_l[Im(BE^{*})+Im(AF^{*})]\Bigg\},\label{pnks}\\
P_T D&=&\frac{\pi\sqrt{\lambda}\hat{m}_l}{4\sqrt{\hat{s}}}
\Bigg\{ 4\hat{s}Re(AB^{*})+\frac{(1-\hat{m}^2_{D_s^*}-\hat{s})}{\hat{m}^2_{D_s^*}}\left[
-Re(BF^{*})+(1-\hat{m}^2_{D_s^*})Re(BG^{*})+\hat{s}Re(BH^{*})\right]\Bigg.\nonumber \\
&&\Bigg.+\frac{\lambda}{\hat{m}^2_{D_s^*}}[Re(CF^{*})-(1-\hat{m}^2_{D_s^*})Re(CG^{*})
-\hat{s}Re(CH^{*})]\Bigg\}. \label{ptks}
\ea

In the family non-universal $Z^\prime$ model, the flavor neutral currents arise even at tree level
owing to non-diagonal chiral coupling matrix. Postulating that the couplings of right-handed quark
flavors with $Z^{\prime}$ boson are diagonal,
the $Z^{\prime}$ part of the effective Hamiltonian for $b\to s l^+ l^-$ transition is described by~\cite{Liu}
\begin{equation}\label{ZPHbsll}
 {\cal H}_{eff}^{Z^{\prime}}(b\to sl^+l^-)=-\frac{2G_F}{\sqrt{2}}
 V_{tb}V^{\ast}_{ts}\Big[-\frac{B_{sb}^{L}B_{ll}^{L}}{V_{tb}V^{\ast}_{ts}}
 (\bar{s}b)_{V-A}(\bar{l}l)_{V-A}-\frac{B_{sb}^{L}B_{ll}^{R}}{V_{tb}V^{\ast}_{ts}}
 (\bar{s}b)_{V-A}(\bar{l}l)_{V+A}\Big]+{\rm h.c.}\,.
\end{equation}

To extract the $Z^{\prime}$ corrections to the Wilson coefficients, one can reformulate Eq.~(\ref{ZPHbsll}) as
\begin{equation}\label{ZPHbsllMo}
 {\cal H}_{eff}^{Z^{\prime}}(b\to sl^+l^-)=-\frac{4G_F}{\sqrt{2}}
 V_{tb}V^{\ast}_{ts}\left[\triangle C_9^{\prime} O_9+\triangle C_{10}^{\prime} O_{10}\right]+{\rm h.c.}\,,
\end{equation}
with
\begin{eqnarray}\label{C910Zp}
 \triangle C_9^{\prime}(M_W)&=&-\frac{g_s^2}{e^2}\frac{B_{sb}^L
 }{V_{ts}^{\ast}V_{tb}} S_{ll}^{LR}\,, \nonumber \\
 \triangle C_{10}^{\prime}(M_W)&=&\frac{g_s^2}{e^2}\frac{B_{sb}^L
 }{V_{ts}^{\ast}V_{tb}} D_{ll}^{LR}\,.
\end{eqnarray}
where $S_{ll}^{LR}=(B_{ll}^{L}+B_{ll}^{R})$, $D_{ll}^{LR}=(B_{ll}^{L}-B_{ll}^{R})$ with $B_{sb}^L$ and
$B_{ll}^{L,R}$ referring to the effective chiral $Z^{\prime}$ couplings to quarks and leptons, respectively.
The off-diagonal element $B_{sb}^L$ contains a new weak phase and can be written as  $|B_{sb}^L|e^{i\phi_s^{L}}$.

When we include the $Z^{\prime}$ contributions with the assumption of no significant RG running effects
between $M_{Z^{\prime}}$ and $M_W$ scales, the Wilson coefficients can be written as
\be
C_{9,10}^{SM}(M_W)\rightarrow C_{9,10}^{SM}(M_W)+\triangle C_{9,10}^{\prime}(M_W)\;.
\ee
After inclusion of the new contributions from $Z^{\prime}$ boson, the RG evolution of the Wilson coefficients down
to low scale is exactly the same as in the SM.

%%%%%%%%%%%%%%%%%%%%%%%%%%%%%%%%%%%%%%%%%%%%
\section{Numerical results}
%%%%%%%%%%%%%%%%%%%%%%%%%%%%%%%%%%%%%%%%%%%%
In this section, we focus on the numerical calculations of the branching ratios,
forward-backward asymmetry and polarization asymmetries for $B_c \rightarrow D_s^* \mu^+ \mu^-$ decay.
The input parameters which are related to our analysis are summarized in Table~\ref{ipp}.
%%%%%%%%%%%%%Table%%%%%%%%%%%%%%%%%%%%%%%%%%%
\begin{table}[t]
 \begin{center}
 \caption{Default values of inputs parameters used in our numerical calculations.}
 \label{ipp}
 \vspace{0.5cm}
 \doublerulesep 0.7pt \tabcolsep 0.1in
 \begin{tabular}{llll} \hline \hline
 $m_{b}=4.8$ GeV, & $m_{B_{c}}=6.28$ GeV, & $m_{D_{s}^{*} }=2.112$ GeV, & $m_{\mu }=0.106$ GeV \\\hline
  $\left|V_{tb}V_{ts}^*\right| = 0.041 $, & $\alpha=1/137$, & $\tau_{B_{c}}=0.46 \times 10^{-12}s$.  & \\
  \hline \hline
 \end{tabular}
 \end{center}
 \end{table}
%%%%%%%%%%%%%%%%%%%%%%%%%%%%%%%%%%%%%%%%%%%%

For the form factors $A_V(s)$, $A_0(s)$, $A_+(s)$, $A_-(s)$, $T_1(s)$,  $T_2(s)$ and $T_3(s)$,
we choose them derived by the three-point QCD sum rules~\cite{formfactor9}, in which the parametrization
of the form factors with respect to $q^2$ are as follows:
\be
F\left( q^{2}\right) =\frac{F\left( 0\right) }{1+ \alpha \hat{s} + \beta \hat{s}^2}.
\label{ff-param}
\ee
where the values of the parameters $F\left( 0\right) $, $\alpha$ and $\beta$ are listed in Table~\ref{formfactor}.
%%%%%%%%%%%%%%%%%%%%%%%%%%%%%%%%%%%%%%%%%%%%%%%%%%%%%%%%%%%
\begin{table}[tbh]
\centering
\caption{$B_{c}\rightarrow D_{s}^{*}$ form factors in the QCD Sum Rules~\cite{formfactor9}.}
\label{formfactor}
\begin{tabular}{cccc}
\hline\hline
 $F(q^{2})$ & $\hspace{2cm}F(0)$ & $\hspace{2cm}\alpha$ & $\hspace{2cm}\beta$ \\ \hline
 $A_{V}\left( q^{2}\right) $ & $\hspace{2.3cm}0.54$ & $%
\hspace{2cm}-1.28$ & $\hspace{2cm}-0.230$ \\ \hline
 $A_{0}(q^{2})$ & $\hspace{2.3cm}0.30$ & $\hspace{2cm}%
-0.13$ & $\hspace{2cm}-0.180$ \\ \hline
 $A_{+}(q^{2})$ & $\hspace{2.3cm}0.36$ & $\hspace{2cm}%
-0.67$ & $\hspace{2cm}-0.066$ \\ \hline
 $A_{-}(q^{2})$ & $\hspace{2cm}-0.57$ & $\hspace{2cm}%
-1.11$ & $\hspace{2cm}-0.140$ \\ \hline
 $T_{1}(q^{2})$ & $\hspace{2.3cm}0.31$ & $\hspace{2cm}%
-1.28$ & $\hspace{2cm}-0.230$ \\ \hline
 $T_{2}(q^{2})$ & $\hspace{2.3cm}0.33$ & $\hspace{2cm}%
-0.10$ & $\hspace{2cm}-0.097$ \\ \hline
 $T_{3}(q^{2})$ & $\hspace{2.3cm}0.29$ & $\hspace{2cm}%
-0.91$ & $\hspace{2.3cm}0.007$ \\ \hline\hline
\end{tabular}
\end{table}
%%%%%%%%%%%%%%%%%%%%%%%%%%%%%%%%%%%%%%%%%%%%%%%%%%%%%%%%%%%
%%%%%%%%%%%%%Table%%%%%%%%%%%%%%%%%%%%%%%%%%%
\begin{table}[t]
 \begin{center}
 \caption{The inputs parameters for the $Z^{\prime}$ couplings~\cite{Chang2,Chang3}. }
 \label{NPPara_value}
 \vspace{0.5cm}
 \doublerulesep 0.7pt \tabcolsep 0.1in
 \begin{tabular}{lcccc} \hline \hline
    & $|B_{sb}^L|(\times10^{-3})$ & $\phi_{s}^L[^{\circ}]$ &$S^{LR}_{\mu\mu}(\times10^{-2})$ & $D^{LR}_{\mu\mu}(\times10^{-2})$\\\hline
 S1 & $1.09\pm0.22$               & $-72\pm7$              &$-2.8\pm3.9$                     & $-6.7\pm2.6$ \\
 S2 & $2.20\pm0.15$               & $-82\pm4$              &$-1.2\pm1.4$                     & $-2.5\pm0.9$ \\
  \hline \hline
 \end{tabular}
 \end{center}
 \end{table}
%%%%%%%%%%%%%%%%%%%%%%%%%%%%%%%%%%%%%%%%%%%%

In the family non-universal $Z^{\prime}$ model, the $Z^{\prime}$ contributions rely on four parameters
$|B_{sb}^L|$, $\phi_{s}^L$, $S^{LR}_{\mu\mu}$ and $D^{LR}_{\mu\mu}$. These parameters have been
constrained from the well measured decays by many groups~\cite{Liu,Chang2,Chang3,CDLv}.
$|B_{sb}^L|$ and $\phi_{s}^L$ have been strictly constrained by $\bar{B}_s-B_s$ mixing,
$B\to\pi K^{(*)}$ and $\rho K$ decays. After taking into account constraints from $\bar{B}_d\to X_s\mu\mu$,
$K\mu\mu$ and $K^{*}\mu\mu$, as well as $B_s\to\mu\mu$ decays, the bounds on $S^{LR}_{\mu\mu}$
and $D^{LR}_{\mu\mu}$ are also obtained. For the sake of convenience, we recollect their numerical
results in Table~\ref{NPPara_value}, with S1 and S2 corresponding to two fitting results of UTfit
Collaboration for $\bar{B}_s-B_s$ mixing~\cite{UTfit}.

Recently, CDF, D0, and LHCb collaborations~\cite{updated1, updated2, updated3} have updated the
CP violation parameter $\phi_s$ in $B_s$ system. These precise measurements will suppress
the magnitude of $b-s-Z'$ coupling by about $10\%$, and have no effect on the new weak
phase $\phi_{s}^L$. However, the weak phase can be constrained by
the data of $B\to\pi K^{(*)}$ and $\rho K$ decays and the results are consistent with
the previous Refs.~\cite{Chang2,Chang3}. Indeed, the quantity that is directly related
to the decay studied here is the product of the couplings of $b-s-Z'$ and $\mu-\mu-Z'$,
and the updated experimental data of $B_s$ mixing have less effect on it.
According to the above analysis, we will adopt the inputs parameters for the $Z^{\prime}$ couplings as
in Table~\ref{NPPara_value} in our theoretical calculation. Meanwhile, we also choose the extreme values of S1 which are
named extreme limit values~(ELV) to show the maximal effects of $Z^{\prime}$ contributions, and the ELV are
\be
\label{ELV}
|B_{sb}^L|=1.31\times10^{-3}\,, \phi_{s}^L=-79^{\circ}\,, S^{LR}_{\mu\mu}= -6.7\times10^{-2}\,, D^{LR}_{\mu\mu}=-9.3\times10^{-2} \,.
\ee

Using the input parameters given above, we obtain the results of the branching ratios
both in the SM and the family non-universal $Z^{\prime}$ model without resonance contributions.
\ba
Br(B_c \to D_s^* \mu^+\mu^-) &=& \left \{
\begin{array}{ll}
2.32^{+0.27}_{-0.26}\times 10^{-7} &  {\rm (SM)}, \\
3.36^{+0.38}_{-0.35}\times 10^{-7} &  {\rm (S1)}, \\
2.80^{+0.32}_{-0.30}\times 10^{-7} &  {\rm (S2)},\\
5.21^{+0.57}_{-0.54}\times 10^{-7} &  {\rm (ELV)}.
\end{array} \right.
\label{br}
\ea
The theoretical errors are induced by the uncertainties of form factors. From the numerical results, one can see that
branching ratio for decay $B_c \to D_s^* \mu^+\mu^-$ is sensitive to the $Z^{\prime}$ contributions.
With respect to the central value of the SM prediction, the new physics contributions in the family non-universal $Z^{\prime}$ model
can provide an enhancement about $45\%$, $21\%$, and $125\%$ for the case of S1, S2, and ELV, respectively.
\begin{figure}[tbp]
\centerline{\epsfxsize=10cm\epsffile{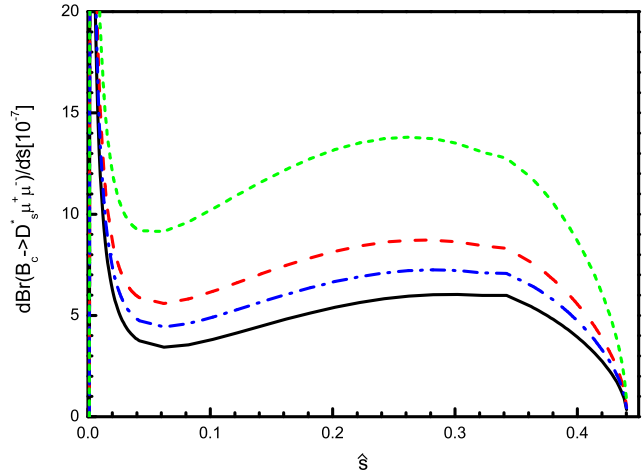}}
\caption{The $\hat{s}$ dependence of the differential decay rates
$dBr(B_c \to D_s^* \mu^+\mu^-)/d\hat{s}$ both in the SM and
the family non-universal $Z^{\prime}$ model. The solid, dashed, dash-dotted,
short-dashed lines show the SM prediction, the
theoretical results of S1, S2, and ELV, respectively. }
\label{fig:dbranching}
\end{figure}

Fig.~\ref{fig:dbranching} shows the $\hat{s}$ dependence of the differential decay rates
for decay $B_c \to D_s^* \mu^+\mu^-$ both in the SM and
the family non-universal $Z^{\prime}$ model using the central values of the input parameters.
The solid line refers to the SM prediction, while the dashed, dash-dotted,
short-dashed curves correspond to the
theoretical results of S1, S2, and ELV, respectively.
The $Z^{\prime}$ enhancements to the differential decay rate are significant in almost the
whole region of $\hat{s}$ and strongly depend on the variation of NP parameters.
\begin{figure}[tbp]
\centerline{\epsfxsize=10cm\epsffile{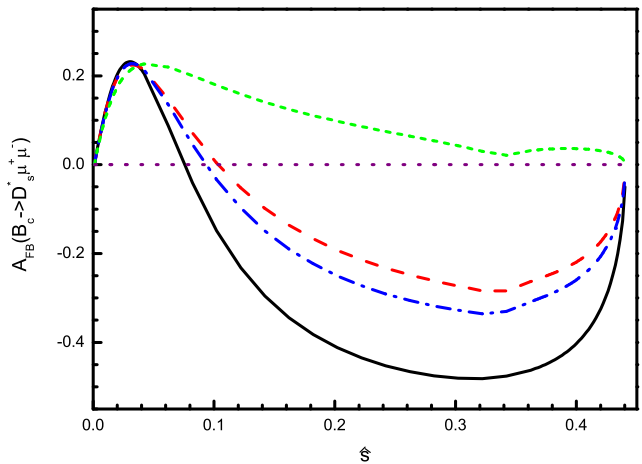}}
\caption{The FBA of decay $B_c \to D_s^* \mu^+\mu^-$ as a function $\hat{s}$ both in the
SM and the family non-universal $Z^{\prime}$ model. } \label{fig:fb}
\end{figure}

The $\hat{s}$ dependence of forward-backward asymmetry for $B_c \to D_s^* \mu^+\mu^-$ decay is
presented in Fig.~\ref{fig:fb}. Compared to the SM results, when including the NP effects from $Z^{\prime}$ boson,
the FBA can be increased by about $47\%$, $38\%$, and $110\%$ at most in S1, S2, and ELV, respectively.
It is easy to see that the zero crossing in $A_{FB}(B_c \to D_s^* \mu^+\mu^-)$ also exists and $Z^{\prime}$
corrections can shift $\hat{s}_0=0.075$ in the SM to $\hat{s}_0=0.104$ in S1, and $\hat{s}_0=0.093$ in S2, respectively.
As for the case of ELV, the $Z^{\prime}$ effects on $A_{FB}(B_c \to D_s^* \mu^+\mu^-)$ are more significant and
can lead zero crossing to vanish.
\begin{figure}[tbp]
\centerline{\epsfxsize=10cm\epsffile{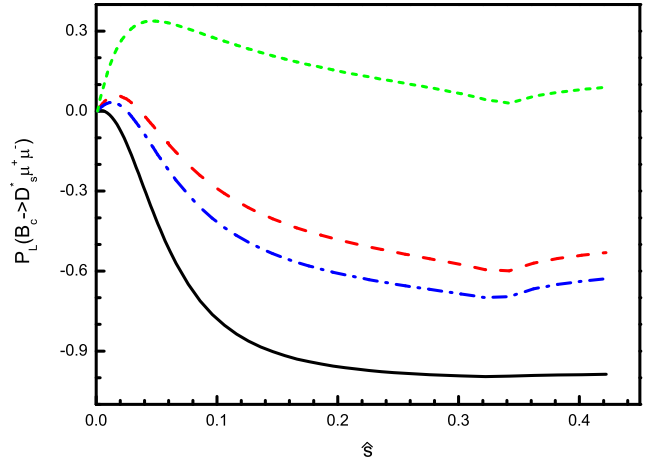}}
\caption{The longitudinal lepton polarization asymmetry of decay $B_c \to D_s^* \mu^+\mu^-$ as a function $\hat{s}$ both in the
SM and the family non-universal $Z^{\prime}$ model. } \label{fig:PL}
\end{figure}

In Fig.~\ref{fig:PL}, we plot the longitudinal lepton polarization asymmetry of decay $B_c \to D_s^* \mu^+\mu^-$ as
a function $\hat{s}$ both in the SM and the family non-universal $Z^{\prime}$ model.
After inclusion of the $Z^{\prime}$ contributions, there are also apparent deviations in the values
of the $P_L(B_c \to D_s^* \mu^+\mu^-)$ for all the cases in $Z^{\prime}$ model from that of the SM predictions.
When $\hat{s}>0.08$, the value of the longitudinal polarization asymmetry can be changed from $-1$ in the SM to $-0.5$
in S1, and $-0.6$ in S2, respectively. In the extreme case, the $Z^{\prime}$ effects
could flip the sign of the SM predictions when $\hat{s}>0.006$ and the theoretical values
might be close to zero in large momentum region.
\begin{figure}[tbp]
\centerline{\epsfxsize=10cm\epsffile{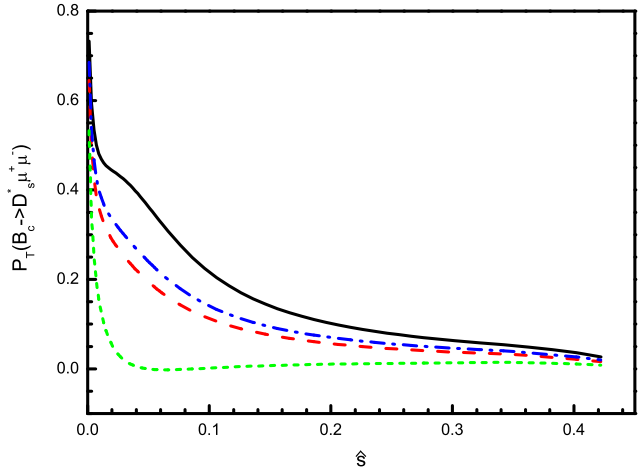}}
\caption{The transverse lepton polarization asymmetry of decay $B_c \to D_s^* \mu^+\mu^-$ as a function $\hat{s}$ both in the
SM and the family non-universal $Z^{\prime}$ model.} \label{fig:PT}
\end{figure}
%%%%%%%%%%%%%%%%%%%%%%%%%%%%%%%%%%%%%%%%%%%

The transverse lepton polarization asymmetry of decay $B_c \to D_s^* \mu^+\mu^-$ as a function $\hat{s}$ both in the
SM and the family non-universal $Z^{\prime}$ model is given in Fig.~\ref{fig:PT}.
The new physics corrections from $Z^{\prime}$ boson are small, and will decrease the SM prediction
about $25\%$ for the cases of S1 and S2 in low $\hat{s}$ region. However, for the case of ELV, the decrease
could be rather large and reach $100\%$ of the SM predictions. In addition, the sign of $P_T$ will be changed
in low momentum region and its values approach to zero when $\hat{s}>0.037$.

\section{Summary}
In this paper, we calculated the $Z^{\prime}$ contributions to the branching ratio,
forward-backward asymmetry and polarization asymmetries for $B_c \to D_s^* \mu^+\mu^-$ decay
in the family non-universal $Z^{\prime}$ model by employing the effective
Hamiltonian with the form factors calculated in the three-point QCD sum rules.

In Section~2, we presented the theoretical framework of $b\to s l^+ l^-$ transition including the effective
Hamiltonian, matrix element and the physical observables. In Section~3, we showed the numerical results of the observables and
made phenomenological analysis for $B_c \to D^*_s \mu^+\mu^-$ decay in the SM and the family non-universal $Z^{\prime}$ model.

As expected, the $Z^{\prime}$ contributions to the observables for $B_c \to D_s^* \mu^+\mu^-$ decay
could be significant in size. From the numerical results, we found that:

\begin{itemize}
\item With respect to the SM prediction, the $Z^{\prime}$ contributions to the differential decay rates are significant in almost the
whole region of $\hat{s}$ and strongly depend on the variation of NP parameters.

\item The new physics enhancements to FBA could be large, and reach $47\%$, $38\%$, and $110\%$ at most in S1, S2, and ELV, respectively.
The zero crossing could be shifted from $\hat{s}_0=0.075$ in the SM to $\hat{s}_0=0.104$ in S1, and
$\hat{s}_0=0.093$ in S2, respectively.
As for the case of ELV, the $Z^{\prime}$ effects could lead zero crossing to vanish.

\item The values of $P_L$ deviated apparently from that of the SM predictions for all the cases
in $Z^{\prime}$ model. In high $\hat{s}$ region, the values of $P_L$ could be changed from $-1$ in the SM to $-0.5$
in S1, $-0.6$ in S2, and 0 in ELV, respectively.

\item The new physics corrections to $P_T$ would decrease the SM prediction
about $25\%$ for the cases of S1 and S2 in low $\hat{s}$ region. However, for the case of ELV, the decrease
could be rather large and reach $100\%$ of the SM predictions.
\end{itemize}

\section*{Acknowledgments}
One of the authors Lin-Xia L\"u would like to thank Prof. Zhen-jun Xiao for his valuable help.
The work is supported by the National Natural Science Foundation of China
under Grant No.~10947020 and 11147004, and Natural Science Foundation of
Henan Province under Grant No.~112300410188.

%%%%%%%%%%%%%%%%%%%%%%%%
\begin{appendix}
%%%%%%%%%%%%%%%%%%%%%%%%
\section*{Appendix A: Auxiliary functions}
The auxiliary functions are given as follows~\cite{Ali-prd61,liwenjun}:
\ba
A(\hat{s})&=&\frac{2}{1+\hat{m}_{D_s^*}} \widetilde
{C}_9^{eff}(\hat{s})A_V(\hat{s})
+\frac{4\hat{m}_b}{\hat{s}}\widetilde C_7^{eff} T_1(\hat{s}),\label{ahs}\\
B(\hat{s})&=&(1+\hat{m}_{D_s^*})\widetilde
{C}_9^{eff}(\hat{s})A_0(\hat{s})
+ \frac{2\hat{m}_b}{\hat{s}}(1-\hat{m}^2_{D_s^*})\widetilde C_7^{eff}T_2(\hat{s}),
\label{bhs}\\
C(\hat{s})&=&\frac{1}{1+\hat{m}_{D_s^*}}\widetilde
{C}_9^{eff}(\hat{s})A_+(\hat{s})
+\frac{2\hat{m}_b}{1-\hat{m}^2_{D_s^*}}\widetilde
C_7^{eff}\left(T_3(\hat{s})+
\frac{1-\hat{m}^2_{D_s^*}}{\hat{s}}T_2(\hat{s})\right),\label{chs}\\
E(\hat{s})&=&\frac{2}{1+\hat{m}_{D_s^*}}
\widetilde C_{10}^{eff}A_V(\hat{s}),\label{ehs}\\
F(\hat{s})&=&(1+\hat{m}_{D_s^*})\widetilde C_{10}^{eff}A_0(\hat{s}),\label{fhs}\\
G(\hat{s})&=&\frac{1}{1+\hat{m}_{D_s^*}}\widetilde C_{10}^{eff}A_+(\hat{s}),\label{ghs}\\
H(\hat{s})&=&\frac{2\hat{m}_{D_s^*}}{\hat{s}}\widetilde C_{10}^{eff}A_-(\hat{s}),\label{hhs}
\ea
\end{appendix}

%%%%%%%%%%%%%%%%%%%%%%%%%%%%%%%%%%%%%%%%%%%%%%%%%%%%%%%%

\end{document}